\documentclass[10pt]{article}
\usepackage{chicago,doublespace}

\begin{document}

\begin{singlespace}
\title{No Information Can Be Conveyed By Certain Events:\\
The Case of the Clever Widows of Fornicalia\\And the Stobon Oracle}

\author{Anand Venkataraman and Ray Kemp\\
Computer Science, IIST\\Massey University\\
New Zealand\\mailto:\{A.Raman,R.Kemp\}@massey.ac.nz}

\maketitle

\begin{abstract}

\begin{quote}
In this short article, we look at an old logical puzzle, its
solution and proof and discuss some interesting aspects
concerning its representation in a logic programming
language like Prolog.  We also discuss an intriguing
information theoretic aspect of the puzzle.
\end{quote}

\end{abstract}

\end{singlespace}


\section{Introduction}
\label{sec:intro}

\begin{quote}

In a certain village on the remote plains of Fornicalia
there exist some men who are having affairs with the wives
of other men.  Now there is a gruesome custom in this
village which requires a woman to kill her husband the
morning after she discovers that he is having an affair with
another woman.  It also happens that every woman knows
whether every other man is having an affair or not except
her own husband.\footnote{A further important fact that is
often not emphasized in most renditions of this puzzle is
that not only does each woman know the character of every
other man in the village except her own husband, she also
knows this to be true of all the women in the village.  It
is also assumed that the women in this village possess the
intelligence required to make logical inferences and that
they recognize this ability in each other.}  So life in this
village goes on peacefully since no woman can know for sure
that her own husband is cheating on her.  Unfortunately, an
Oracle from the pure and untainted shores of distant Stobon
visits the village one day and proclaims that at least one
man in this village is having an affair.  What happens after
this?

\end{quote}

Readers are urged to think of a solution themselves before
proceeding further.  We first present the solution and its
proof and finally go on to discuss the information theoretic
aspects of this remarkable problem.\footnote{A reader
has remarked that ``In older kinder times, descriptions of
problems of this sort were less grisly.  A person would have
a black spot painted on his forehead or not.  No mirrors
exist in the community, so each person would know if another
had a black spot, but he could not {\em directly} know about
himself.''}  The first two sections of this short note will
be of use as a reading for undergraduate Computer Science
students learning its foundations.  It's final two sections
are independent of each other and may be read in any order.
Section~\ref{sec:prolog} is a nice Prolog representation of
this proof that can be used to visualise and feel
comfortable with it for those of us, including from time to
time, the authors, who are still skeptical in spite of
believing the underlying mathematics.
Section~\ref{sec:info-theory} deals with an intriguing
aspect of this problem to do with Information Theory.  It is
likely to be useful to graduate students or computing
professionals being exposed to Information Theory for the
first time.  In any case, the article is bound to be of
interest to anyone with a problem-solving bent of mind.

The answer to the above question is that $n$ killings will
take place on the morning of the $n$th day after the
Oracle's seemingly innocuous but ultimately catastrophic
proclamation, where $n$ is the number of unfaithful men.  If
there are 100 unfaithful men, for example, then nothing will
happen for 99 days, but on the 100th morning, all 100 of the
unfaithful men will be killed!

\section{Proof}
\label{sec:proof}

Let a woman be said to ``know'' that her husband is
unfaithful when she has reason to believe beyond doubt that
he is having an affair.  Since she can never have first-hand
knowledge about the character of her own husband it follows
that her knowledge in this matter is necessarily inferred
fact.  In the absence of this knowledge, she is forced to
subscribe to what we call ``the conservative hypothesis''
that one's own husband is faithful.  If a woman rejects the
conservative hypothesis, then it follows that she has
determined her husband is unfaithful and will thus kill him
the next morning.

Here is the inductive assertion we make on $i$, the number
of unfaithful men in the village:

\begin{quote}
$S(i):$ If there are $i$ unfaithful men in the village, then
$i-1$ mornings after the Oracle's proclamation, each of the
$i$ wives of the $i$ unfaithful men will have conclusive
evidence to dismiss the conservative hypothesis about their
husbands.
\end{quote}

\subsection{Basis}

Consider the case when there is exactly one unfaithful man
$A$ in the village.  Until the Oracle's visit, $A$'s wife
has no reason to believe that her own husband is unfaithful
and will thus continue to subscribe to the conservative
hypothesis.  Immediately after the Oracle's visit, however,
$A$ knows that at least one man in the village is
unfaithful.  Since it cannot be any of the others whom $A$
knows to be faithful, $A$ infers that it must be her own
husband and so kills him on the morning of the first day
following the Oracle's visit.  It has taken $A$'s wife
exactly 0 mornings to determine whether her husband was
faithful or not.  The basis is therefore true.

\subsection{Inductive step}

Assume that our assertion $S(i)$ is true for all numbers of
unfaithful men less than $n$, i.e. for $0 < i < n$.  Now
suppose that the village has $n$ unfaithful men.  We prove
by complete induction on $n$ that $S(i)$ still holds.

When $i=n$, there are two possible implications that could
result from the conservative hypotheses of the Fornicalian
women:

\begin{enumerate}
  \item $n-1$ men in the village are unfaithful
  \item $n$ men in the village are unfaithful
\end{enumerate}

Statement 1 is the implication of the conservative
hypothesis held by the wives of each of the $n$ unfaithful
men, since they know $n-1$ other unfaithful men and have no
reason to believe their own husbands are unfaithful.
Statement 2 is the implication of the conservative
hypothesis held by the rest of the women in the village.

By the inductive hypothesis, the believers of statement 1
will expect $n-1$ killings to take place on the morning of
the $n-1$st day after the Oracle's visit.  However, this
doesn't happen since each of the $n$ women will expect to
see the husbands of the other $n-1$ women dead and not their
own.  Therefore, on the morning when this expected killing
doesn't take place, they are forced to reject their
conservative hypothesis and conclude there must be more than
$n-1$ unfaithful men in the village.  Since each of the
women know that exactly $n-1$ other men are unfaithful, each
concludes that the only possibility is that $n$ men must be
unfaithful and that the $n$th unfaithful man must be her own
husband.  She thus kills him on the morning of the $n$th
day.  It has taken the $n$ wives of the $n$ unfaithful men
exactly $n-1$ mornings to determine that their husbands were
unfaithful.  This completes the inductive step and so $S(i)$
is true for all $n$.

\section{The proof in Prolog}
\label{sec:prolog}

Modelling the puzzle in Prolog is an interesting exercise.
While it tackles the problem from a different angle to shed
further light on it, it also tests how effective a logic
programming language would be for representing and solving
problems of this kind.

\subsection{Prolog preliminaries}

The main assertion, which we will call $s(N)$, is that when
there are $N$ unfaithful husbands then the $N$ wives will
discover this on day $N-1$.  This can be represented by a
Prolog predicate $s(N)$ which will succeed if the assertion
is true and will fail otherwise.  $N$, of course, must be a
positive number so a first attempt at specifying our goal
might be this:
\begin{singlespace}
\begin{verbatim}
s(N), N > 0.
\end{verbatim}
\end{singlespace}

However, this will not do on two counts. First, since
Standard Prolog \cite{Deransart:PTS96} does not have either
implicit or explicit typing of variables there is no
guarantee that $N$ will be a number during a call. Secondly,
Prolog insists that $N$ must have been instantiated to a
constant value by the time the call $N > 0$ takes place.  We
don't wish to be this restrictive since we need to check
that it is satisfied for all positive $N$, so will want to be
able to make the call with $N$ uninstantiated.

A way of solving both problems at once is to represent
numbers in structured form (see, for example,
\shortciteN{Sterling:TAP94}). We will slightly amend their
representation, and denote zero by 0, 1 by 0+1, 2 by 0+1+1,
etc.

To recognise numbers in this form we set up a predicate,
{\tt natural\_number(N)} which succeeds if $N$ is a
number in this form and fails otherwise. A recursive
definition of this predicate is given below:
\begin{singlespace}
\begin{verbatim}
natural_number(N):-var(N),!.
natural_number(0).
natural_number(N+1):- natural_number(N).
\end{verbatim}
\end{singlespace}

Notice that we allow any uninstantiated variable to denote a
number which is what is needed for the general case. The two
operations that we need to be able to carry out on these
numbers are checking for equality, and checking for one
being less than another.

We cannot use the standard unifier operator, $=$, for
checking equality since a call such as $N = N+1$ will send
most Prolog systems to sleep.  This is because the variable
on the left hand side also occurs in the expression on the
right hand side.  To get around this problem, we use the
Standard Prolog predicate {\tt unify\_with\_occurs\_check}
to ensure this kind of comparison fails and then create an
operator {\tt eq} for checking equality of numbers thus:
\begin{singlespace}
\begin{verbatim}
:-op(700,xfx,eq).

0 eq 0.
A+1 eq B+1 :- unify_with_occurs_check(A,B), 
              natural_number(A), natural_number(B).
\end{verbatim}
\end{singlespace}

An operator, {\tt lt}, for checking whether one number is
smaller than another can similarly be set up:
\begin{singlespace}
\begin{verbatim}
:- op(700, xfx, lt).

0 lt K+1 :- natural_number(K).
A+1 lt B+1 :-  \+ (A eq B), A lt B.
\end{verbatim}
\end{singlespace}

``$\backslash+$'' is the Standard Prolog symbol for `not
provable'.  Note that we need to check for $A$ being equal
to $B$.  Otherwise the test will loop in this situation.

\subsection{Proof representation}

Moving on now to how the inductive assertion, $s(N)$, can be
expressed in terms of the problem, we first create a
predicate, {\tt reject\_conservative\_hypothesis},
which indicates the number of days that it will take the
betrayed women of Fornicalia to abandon their conservative
hypothesis.  This can be written as follows:
\begin{singlespace}
\begin{verbatim}
reject_conservative_hypothesis(I, N, Day)
\end{verbatim}
\end{singlespace}

\noindent where $I$ is the inductive counter, $N$ denotes
the test case, and $Day$ is the day on which the women can
work out the truth. Now we express $s(N)$ in terms of {\tt
reject\_conservative\_hypothesis}:
\begin{singlespace}
\begin{verbatim}
s(N):- reject_conservative_hypothesis(N, N, Day), N eq Day+1.
\end{verbatim}
\end{singlespace}

The above statement says that if there are $N$ unfaithful
husbands then the wives will discover this on day number
$N-1$. Note that, as it happens, we do not have to stipulate
$N > 0$ since this is implicit in the term $N=Day+1$.

By the inductive assumption, if the number of betrayed
women, $I$, is less than $N$ then they will discover this on
day $I-1$. This can be written:
\begin{singlespace}
\begin{verbatim}
reject_conservative_hypothesis(I, N, Day) :-  I lt N, I eq Day+1.
\end{verbatim}
\end{singlespace}

Then, of course, the killings take place.  We introduce a
predicate called {\tt kill} to indicate when this occurs:
\begin{singlespace}
\begin{verbatim}
kill(I, Day)
\end{verbatim}
\end{singlespace}

This happens, for all $I$, on the day after the women learn
the truth. So we have:
\begin{singlespace}
\begin{verbatim}
kill(I, Day+1) :- reject_conservative_hypothesis(I, N, Day).
\end{verbatim}
\end{singlespace}

Now we need to determine when the betrayed women find out in
the case when $I = N$.  The $N$ women in this case subscribe
to the conservative hypothesis and therefore believe there
are $N-1$ unfaithful husbands. They will, therefore, expect
these husbands to be found out on day $N-2$, and to be
killed on day $N-1$. If this is so then the call:
\begin{singlespace}
\begin{verbatim}
?- kill(N, Day).
\end{verbatim}
\end{singlespace}

\noindent where $N = Day + 1$ will succeed. If it doesn't
then they will know their own husbands are unfaithful. We
can express this in the Prolog clause:
\begin{singlespace}
\begin{verbatim}
reject_conservative_hypothesis(N, N, Day) :-  N eq Day+1,
                                              \+ kill(N, Day).
\end{verbatim}
\end{singlespace}

Finally, we have to check that the basis is
satisfied. The simplest way of doing this is to make the
call:
\begin{singlespace}
\begin{verbatim}
s(0+1)
\end{verbatim}
\end{singlespace}

\noindent to check whether it succeeds. Not only this, but we must
ensure it does not use the inductive rule.  A trace of the program,
run under SICStus Prolog \shortcite{Andersson:SPUM93} is shown in
Figure~\ref{fig:progtrace-base}.  It verifies that, indeed, the proof
does not use the inductive rule and so is satisfied.

\begin{figure}[htb]
\begin{center}
\begin{singlespace}
\begin{verbatim}
| ?- s(0+1).
Call: s(0+1) 
   Call: reject_conservative_hypothesis(0+1,0+1,Day) 
      Call: 0+1 lt 0+1 
         Call: 0+1 eq 0+1 
         Exit: 0+1 eq 0+1 
      Fail: 0+1 lt 0+1 
      Call: 0+1 eq Day+1 
      Exit: 0+1 eq 0+1 
         Call: \+ kill(0+1,0) 
            Call: kill(0+1,0) 
            Fail: kill(0+1,0) 
         Exit: \+ kill(0+1,0) 
   Exit: reject_conservative_hypothesis(0+1,0+1,0) 
   Call: 0+1 eq 0+1 
   Exit: 0+1 eq 0+1 
Exit: s(0+1) 
\end{verbatim}
\end{singlespace}
\caption{A program trace of the basis of the proof in SICStus Prolog}
\label{fig:progtrace-base}
\end{center}
\end{figure}

If we now make a call with any legal value of $N$, ($N=0+1+1+1$,
for example), then $S$ will succeed. Indeed, if we make the
general call:
\begin{singlespace}
\begin{verbatim}
?- s(N).
\end{verbatim}
\end{singlespace}

\noindent then the solution will be:
\begin{singlespace}
\begin{verbatim}
N = _ + 1
\end{verbatim}
\end{singlespace}

\noindent where ``{\tt \_}'' is the anonymous variable denoting any value.
Figure~\ref{fig:progtrace-gen} shows a program trace for this general
case.  Thus we can infer that the assertion is true for all values of
$N$.

\begin{figure}[htb]
\begin{center}
\begin{singlespace}
\begin{verbatim}
| ?- s(N).
Call: s(N) 
   Call: reject_conservative_hypothesis(N,N,Day) 
      Call: N lt N 
      Fail: N lt N 
      Call: N eq Day+1 
      Exit: Day+1 eq Day+1 
      Call: \+ kill(Day+1,Day) 
         Call: kill(Day+1,Day) 
            Call: reject_conservative_hypothesis(Day+1+1,N,Day) 
               Call: Day+1+1 lt N   
               Fail: Day+1+1 lt N 
               Call: Day+1+1 eq Day+1 
               Fail: Day+1+1 eq Day+1 
            Fail: reject_conservative_hypothesis(Day+1+1,N,Day) 
         Fail: kill(Day+1,Day) 
      Exit: \+ kill(Day+1,Day) 
   Exit: reject_conservative_hypothesis(Day+1,Day+1,Day) 
   Call: Day+1 eq Day+1 
   Exit: Day+1 eq Day+1 
Exit: s(Day+1) 
\end{verbatim}
\end{singlespace}
\caption{A program trace of the general case of proof in SICStus Prolog}
\label{fig:progtrace-gen}
\end{center}
\end{figure}

Also, if we wish to find out which day the unfaithful
husbands are discovered then we can make the call
\begin{singlespace}
\begin{verbatim}
?- reject_conservative_hypothesis(N, N, Day).
\end{verbatim}
\end{singlespace}

\noindent which gives the general solution:
\begin{singlespace}
\begin{verbatim}
N = Day + 1 ,
Day = _
\end{verbatim}
\end{singlespace}

Clearly it is possible to represent the problem fairly
directly in Prolog and also to use the language's deductive
mechanism to verify the solution.  However, the
representation and manipulation of numbers is rather clumsy
and the program produced is not purely declarative.  It may
be that using a more sophisticated logic programming
language such as Mercury \shortcite{Somogyi:MAE95} would
produce a cleaner solution.  A more significant limitation
worth mentioning is that the program only confirms the
solution --- the insight enabling the problem to be solved
in the first place has been provided by a human and is built
into the program.

\section{Information Content of the Proclamation}
\label{sec:info-theory}

Surprisingly, the mathematical proof presented above
requires the Oracle to very much be an essential part of
this situation.  If not for the Oracle, the basis
fails.  Yet, it seems on a first reading that the Oracle's
statement contains no useful information in a village that
has more than one unfaithful man.  Consider the village that
has two unfaithful men, for example.  Every woman in the
village knows at least one unfaithful man.  Thus the
Oracle's proclamation is at best, one could say, ``stating
the bleeding obvious.''  A foundational result in
Information Theory due to
\shortciteN[p.82--83]{Shannon:MTC49} shows that when the term
{\em information\/} is sensibly defined, the information
content of a symbol is equal to the negative logarithm of
its probability.  We also point the reader at another
seminal work, \shortciteN{Hamming:CIT80}, where the former
interesting result is more accessibly discussed at length,
and at the URL:
http://cm.bell-labs.com/cm/ms/what/shannonday/ from where a
copy of Shannon's original paper can be downloaded.  This
latter document, however, doesn't contain Weaver's lucid
discussion of the significance of Shannon's results.

Claiming that the information content of a symbol is related
to the negative logarithm of its probability is tantamount
to claiming that no useful information can be conveyed by a
certain event.  The more uncertain and therefore surprising
an event is, the more information it contains.  If we told
you, for example, that the Sun rose in the East this
morning, you will hopefully not benefit much from this fact
by virtue of your already attaching a very high probability
to this event.  However, if we were to tell you instead that
the Sun happened to rise in the West, that information,
assuming it is true, will be of utmost interest and use to
you.  In general the information content in an event $i$ is
$-\log p_i$ where $p_i$ is the probability of the event.

The term {\em information\/} is used somewhat differently in
logic and in information theory.  In the latter we are able
to specify a single real number that uniquely characterises
the information content of a source, but not so in the
former.  Yet, it turns out that Shannon's results are just
as profoundly valid in logic or epistemology as they are
elsewhere.  For instance, each observer will attach a
subjective probability to a given statement being true.  And
to this observer, the amount of information contained in an
event that exposes the objective truth value of that
statement is in fact equal to the negative logarithm of this
subjective probability.  So at the very least, if we know
whether an event was certain or not, we can associate either
zero information content with it or not.  Naturally, we
would expect an event void of information to have no effect
upon its recipient and thus only expect effects to be caused
by uncertain events, however small this uncertainty is.  The
point that is relevant to our discussion here is that while
we are unable to say precisely how much subjective
information is contained in an uncertain event for each
person, we are able to say with certainty that a sure event
contains zero information for each person concerned.

Now suppose again that there are at least two unfaithful men
in the village.  Every wife in the village knows at least
one unfaithful man and so the subjective probability she
will assign to the event that there is at least one
unfaithful man in the village is 1.0.  It follows,
therefore, that the information contained for her in the
oracular statement that claims this certain event is $-\log
1 = 0$.  So how can this event, seemingly void of
information, cause such a catastrophic result?  Let us
rephrase the problem to make it even more explicit.  {\bf
The Oracle has not told the women of Fornicalia anything new
that they didn't already know.  So why was its proclamation
critical in determining the subsequent course of events in
the village?}  Pondering this question, more than any other,
promises to be most instructive for someone just being
introduced to the subtleties of Information Theory.  Again,
we invite the reader at this point to contemplate why this
is so before proceeding further.

\subsection{Why the Oracle is essential}

It turns out that the Oracle's proclamation did indeed bear
no useful information for any woman.  And this is precisely
the reason that no killing takes place on the first day
following this event.  But a fact about the proclamation
itself bears useful information for just the wives of the
two unfaithful men in the village.  In particular, it is the
{\em meta-fact\/} that {\em the oracular proclamation had
zero information for every woman in the village}.  In other
words, although every woman knew that the village had at
least one unfaithful man, not every woman knew that ``every
woman knew that the village had at least one unfaithful
man''.  Consequently, these women will attach a non-unitary
probability to this event and will thus find it useful.
Let's review the case when there are exactly two unfaithful
men in the village.  Call their wives $A$ and $B$.  Every
woman in the village knows at least one unfaithful man.  But
not every woman knows that every woman knows at least one
unfaithful man.  To be precise, $A$ and $B$ know only one
unfaithful man in the village and $A$ does not know that $B$
knows that there exists at least one unfaithful man and
vice-versa.  Thus the fact that $B$ does know this comes as
a surprise to $A$ after the first morning.  In general, one
can make the following assertion for any number, $i > 0$, of
unfaithful men in the village:
\begin{quote}
$T(i):$ If there are $i$ unfaithful men in the village, then
the wives of these $i$ men don't know (that every woman
knows)$^{i-1}$ that there exists at least one unfaithful man
in the village.
\end{quote}
where the superscripted index denotes $i-1$ repetitions of
the parenthesized phrase.  It is easy to prove $T(i)$ by
induction along the lines of our previous proof in the first
section.  Thus those women who don't know the above fact
will attach a subjective probability $p < 1$ to it's being
true and consequently find it of informative value.  So the
Oracle is an integral part of the situation after all.  Had
it not visited the village, the catastrophe wouldn't have
been triggered off and the village would have continued to
exist as a harmonious society in which no woman can
conclusively nail down her husband.  However, that being not
the case, thus ends the lamentable tale of the Stobon Oracle
and the clever widows of Fornicalia.\footnote{Imagine the
even more lamentable case when there is exactly one
unfaithful man and his wife also happens to be the only free
thinking secret rebel in the village who doesn't believe in
its gruesome custom.  Suppose she demonstrates her
recalcitrance by not killing her husband on the morning of
the first day after the Oracle's proclamation.  What happens
then?}  The men's lives wouldn't have been lost in vain if
their story has inspired in new students a deep and lasting
love for Information Theory.

\bibliography{village}
\bibliographystyle{chicagoa}


\end{document}